\documentclass[prd,twocolumn]{revtex4}
\usepackage{graphicx}
\begin{document}

\newcommand{\be}{\begin{equation}}
\newcommand{\ee}{\end{equation}}
\newcommand{\bea}{\begin{eqnarray}}
\newcommand{\eea}{\end{eqnarray}}
\title{\bf Higgs Decays and Brane Gravi-vectors}
\author{T.E. Clark}
\email{clark@physics.purdue.edu}
\affiliation{Department of Physics,
Purdue University,
West Lafayette, IN 47907-2036, U.S.A.}
\author{Boyang Liu}
\email{liu115@purdue.edu}
\affiliation{Department of Physics,
Purdue University,
West Lafayette, IN 47907-2036, U.S.A.}
\author{S.T. Love}
\email{love@physics.purdue.edu}
\affiliation{Department of Physics,
Purdue University,
West Lafayette, IN 47907-2036, U.S.A.}
\author{T. ter Veldhuis}
\email{terveldhuis@macalester.edu}
\affiliation{Department of Physics \& Astronomy,
Macalester College,
Saint Paul, MN 55105-1899, U.S.A.}
\author{C. Xiong}
\email{xiong@purdue.edu}
\affiliation{Department of Physics,
Purdue University,
West Lafayette, IN 47907-2036, U.S.A.}
\begin{abstract}
Higgs boson decays in flexible brane world models with stable, massive gravi-vectors are considered. Such vectors couple bilinearly to the Standard Model fields through either the Standard Model energy-momentum tensor, the weak hypercharge field strength or the Higgs scalar. The role of the coupling involving the extrinsic curvature is highlighted. It is found that within the presently allowed parameter space, the decay rate of the Higgs into two gravi-vectors (which would appear as an invisible Higgs decay) can be comparable to the rate for any of the Standard Model decay modes.
\end{abstract}

\maketitle

Various theoretical extensions of theories of gravity include vector particles. In particular, such gravi-vectors\cite{local} appear in flexible brane world models in which a four dimensional spacetime is embedded in a higher dimensional spacetime thus breaking the extra dimensional spatial translation symmetries\cite{coset}. When these symmetries are made local thereby  including  higher dimensional gravitational interactions, the erstwhile Nambu-Goldstone scalar degrees of freedom\cite{branon} associated with the higher dimensional spatial translation symmetry breakdown become the longitudinal components of the now massive vector particles, $X_i^\mu$. For  $N\ge 2$ additional compactified isotropic spatial dimensions, these vectors, which are completely neutral under the Standard Model gauge group, carry an additional $SO(N)$ quantum number, labelled by $i=1,...,N$, which reflects the isometry of the co-dimensional space when the 4-dimensional brane is embedded in the larger dimensional spacetime. On the other hand, all  Standard Model particles are $SO(N)$ singlets.  Consequently, $SO(N)$ invariant interactions of these  vector Proca fields require them to appear in pairs and they are thus massive, stable physical degrees of freedom. For a single codimension, $N=1$, the vector is stable provided there is a unbroken parity with respect to the extra dimension under which the brane vector is odd. Being stable, the vectors are also candidates for the dark matter of the universe and are thus subject to the appropriate constaints\cite{phenom}.

The strength of the vector interactions is governed by the ratio $\frac{M_X^2}{F_X^4}$ where $M_X$ is the vector mass whose non-zero value is model dependent and $F_X^4$ the brane tension. In the flexible brane limit where the brane tension is much smaller than the $D$-dimensional Planck scale, the Kaluza-Klein modes of higher dimensional
gravity decouple from the Standard Model particles and thus we can focus attention on the coupling of the brane-vectors to the Standard Model. The leading four-dimensional couplings of these vectors to the Standard Model is given by\cite{phenom} 
\be
{S}_{X-SM}=\int d^4x [{\cal L}_{TXX} +{\cal L}_{BXX}+{\cal L}_{HXX}] ~.
\label{Eff}
\ee
Here the effective Lagrangian
\be
{\cal L}_{TXX}=\frac{1}{2}\frac{M_X^2}{F_X^4}X^\mu_i T^{\rm SM}_{\mu\nu} X^\nu_i 
\label{emtensor}
\ee
details the coupling of the induced metric on the brane to the Standard Model symmetric energy momentum tensor $T^{\rm SM}_{\mu\nu}$. The other two interactions involve the extrinsic curvature and vector field strength tensors. The extrinsic curvature tensor 
\be
K^{\mu\nu}_i=\frac{1}{2}(\partial^\mu X^\nu_i+\partial^\nu X^\mu_i)+...
\ee
measures the curvature of the embedded  brane relative to enveloping $D$-dimensional geometry, while the vector field strength is given by
\be 
X_i^{\mu\nu}=\partial^\mu X_i^\nu-\partial^\nu X_i^\mu ~.
\ee
The product of the extrinsic curvature tensor and field strength couple to the Standard Model singlet weak hypercharge field strength  $B_{\mu\nu}=\partial_\mu B_\nu-\partial_\nu B_\mu$ with $B_\mu =\cos\theta_W A_\mu -\sin\theta_W Z_\mu$ as 
\be
{\cal L}_{BXX}=\frac{M_X^2}{2F_X^4}[\kappa_1 B_{\mu\nu}+\kappa_2 \tilde{B}_{\mu\nu}] X_i^{\mu\rho}K_{i\nu}^\rho ~.
\ee
The coefficients $\kappa_1, \kappa_2$ are dimensionless constants of the effective Lagrangian. Finally, there is an invariant 
 coupling to the Standard Model scalar doublet bilinear, $\phi^\dagger \phi$, given by 
\bea
{\cal L}_{HXX}&=&\frac{M_X^2}{2F_X^4}[\lambda_1 K_i^{\mu\nu}K_{i\mu\nu}+\lambda_2X_{i\mu\nu}X_i^{\mu\nu}\cr
&&~~~~~~~~~+\lambda_3X_{i\mu\nu}\tilde{X}_i^{\mu\nu}](\phi^\dagger\phi -\frac{v^2}{2}) ~.
\eea
with $v^2= \frac{M_W^2 \sin^2\theta_W}{\pi \alpha}$. In unitary gauge: $\phi^\dagger\phi -\frac{v^2}{2}= vH+\frac{1}{2}H^2$ with $H$ the  Higgs scalar so that included in ${\cal L}_{HXX}$ is the interaction of the Higgs to two $X$ fields. Here again, the coefficients $\lambda_1, \lambda_2, \lambda_3$ are dimensionless constants  of the effective Lagrangian.

Thus the effective interaction (\ref{Eff}) is characterized by two mass scales, $F_X$ and $M_X$ which we treat as free parameters as well as five model dependent  dimensionless couplings, $\kappa_1, \kappa_2, \lambda_1, \lambda_2,\lambda_3$. Constraints on the $F_X-M_X$ parameter space have been obtained\cite{phenom} using collider data and dark matter limits. For the numerical estimates made in this paper, we shall, for definiteness, use the value $F_X = 250$ GeV which is well within the currently allowed range for $M_X \le 100$ GeV and is appropriate for a Higgs boson with mass less than $200$ GeV. Moreover, for definiteness, we take $N=2$. 

We now turn to consider Higgs boson decays containing a pair of $X$ vectors. We begin with the decay $H\rightarrow XX$ which appears as an invisible Higgs decay. The leading order Feynman graph for this process is displayed in figure 1 
\begin{figure}[h]
\includegraphics[width=2in]{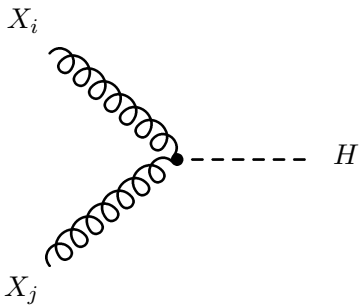}
\caption{Leading order Feynman graph for process $H\rightarrow XX$.}
\end{figure}
where the vertex is secured using the effective Lagrangian ${\cal L}_{HXX}$. Note that ${\cal L}_{TXX}$ does not contribute to the leading order graph since vacuum stability dictates that there is no term linear in $H$ in the Higgs potential. The decay rate is readily computed as 
\bea
&&\Gamma (H\rightarrow XX) =\frac{N\sin^2\theta_W}{32 \pi^2 \alpha}\frac{M_H^7 M_W^2}{F_X^8}\sqrt{1-4\xi^2}\cr
&&~~~~~\biggr[(\frac{\lambda_1}{4} +\lambda_2)^2(1-2\xi^2)^2(\frac{1}{4}-\xi^2 +3\xi^4)\cr
&&~~~~~+(\frac{\lambda_1^2}{16}-\lambda_2^2)\frac{1}{2}(1-2\xi^2)^2(1-4\xi^2)\cr
&&~~~~~+\frac{1}{4}(\frac{\lambda_1}{4}-\lambda_2)^2(1-4\xi^2)^2+8\lambda_3^2 \xi^4(1-4\xi^2)\biggr]
\eea
with $\xi \equiv \frac{M_X}{M_H}$. Figures 2 and 3 display the ratio of this rate to the total Higgs decay rate as computed in the Standard Model for Higgs masses of $120$ GeV and $180$ GeV taking the parameters $\lambda_1=\lambda_2=\lambda_3=1$.
\begin{figure}
\includegraphics[width=3.25in]{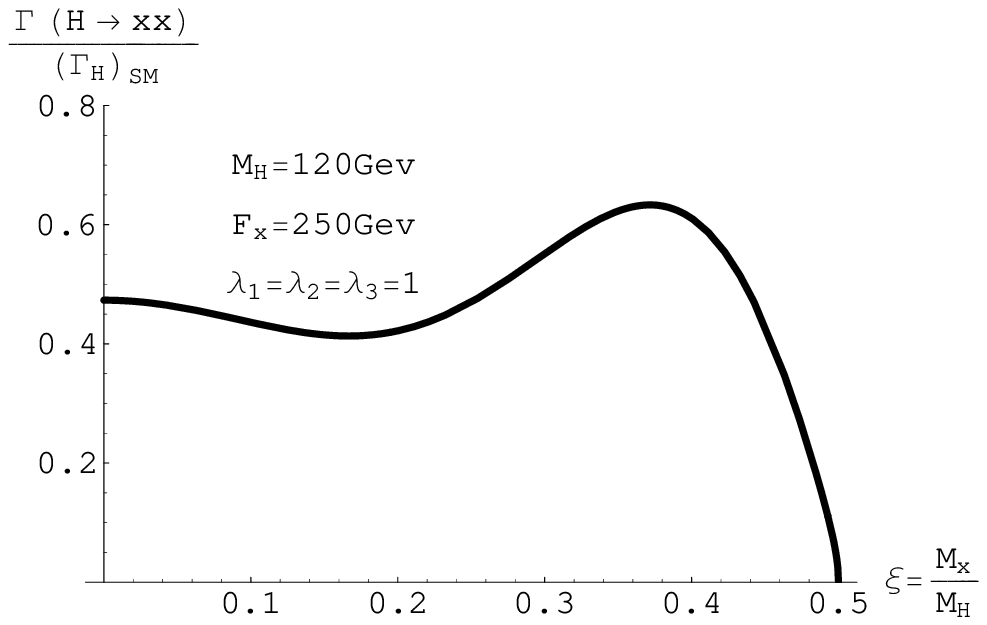}
\caption{Ratio of decay rate for  $H\rightarrow XX$ to the Standard Model Higgs decay rate  as a function of $M_X/M_H$ with $M_H=120$ GeV and $F_X = 250$ GeV.}
\end{figure}
\begin{figure}
\includegraphics[width=3.25in]{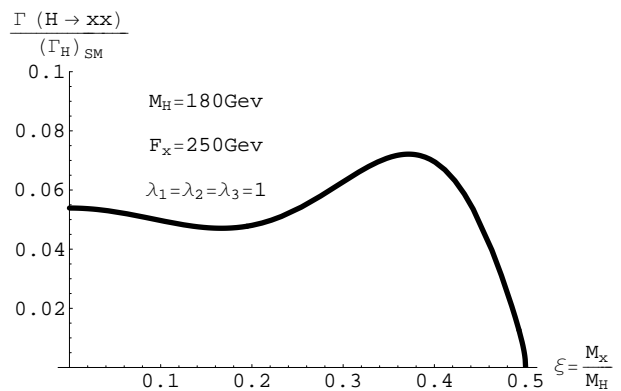}
\caption{Ratio of decay rate for  $H\rightarrow XX$ to the  Standard Model Higgs decay rate  as a function of $M_X/M_H$ with $M_H=180$ GeV and $F_X = 250$ GeV.}
\end{figure}
We see that the rate of Higgs decay to a pair of $X$ vectors for these parameters is quite comparable to the rate for any of the Standard Model decay channels. Note that the $F_X$ dependence is quite steep varying as $1/F_X^8$. Even so, for an $F_X$ value as high as  $500$ GeV, the $120$ GeV Higgs boson will still decay to the $X$ pairs at roughly a rate of $10^{-3}$ that obtained from the Standard Model. 

\begin{figure}
\includegraphics[width=3.25in]{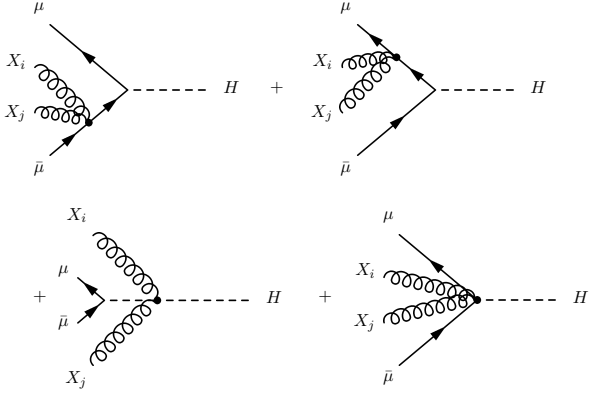}
\caption{Leading order Feynman graphs for the decay $H\rightarrow \mu \bar\mu XX$ mediated using the coupling ${\cal L}_{TXX}$.}
\end{figure}
Since the Higgs decay to $\mu\bar\mu$  is one of considerable focus at the LHC, 
we next consider the decay channel $H\rightarrow \mu\bar\mu XX$ which would appear as $\mu\bar\mu$ plus missing energy. This process can be mediated by either the induced metric coupling ${\cal L}_{TXX}$ 
to the energy-momentum tensor $T^{SM}_{\mu\nu}$ or the coupling ${\cal L}_{BXX}$ to the weak hypercharge. The former interactions appear in the Feynman graphs of figure 4. The amplitude obtained from these graphs varies as $\frac{m_\mu}{F_X^4}$ and is thus, in general, suppressed relative to the amplitude obtained from the graph of figure 5 which depends on the  extrinsic curvature coupling ${\cal L}_{BXX}$  and is independent of $m_\mu$.

\begin{figure}
\includegraphics[width=2in]{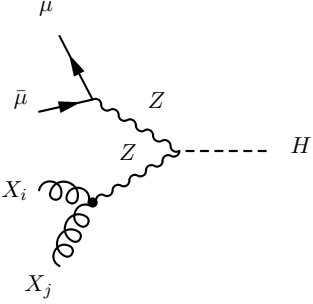}
\caption{Leading order Feynman graph for the decay $H\rightarrow \mu \bar\mu XX$ mediated using the coupling ${\cal L}_{BXX}$.}
\end{figure}

In figures 6-7 is plotted the ratio of the total rate for this process to the Standard Model rate of Higgs decay to $\mu\bar\mu$ for  a Higgs mass of $120$ GeV while taking $\kappa_1=\kappa_2 =1$. The rather sharp peak in figure 6 
\begin{figure}
\includegraphics[width=3.25in]{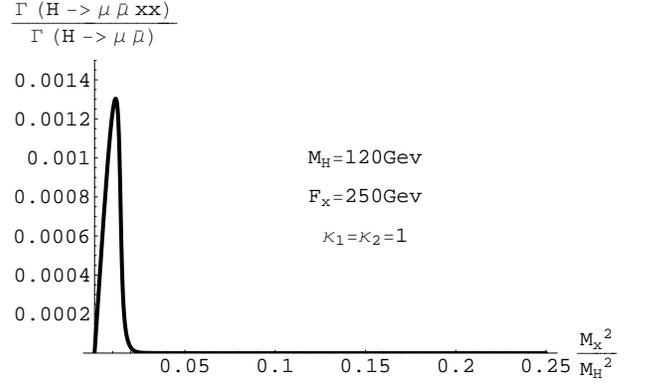}
\caption{Ratio of decay rate for $H\rightarrow \mu\bar\mu XX$ to Standard Model rate for $H\rightarrow \mu \bar\mu$ as a function of $M_X^2/M_H^2$ for $M_H= 120 $ GeV and $F_X= 250$ GeV with the scale on the vertical axis appropriate for displaying the sharp peak at $M_X \simeq 13$ GeV.}
\end{figure}
\begin{figure}
\includegraphics[width=3.25in]{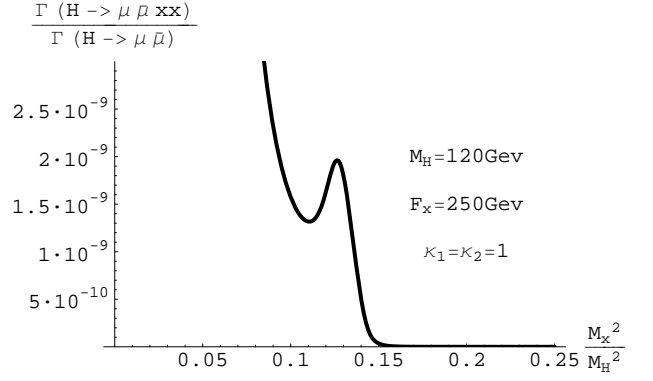}
\caption{Ratio of decay rate for $H\rightarrow \mu\bar\mu XX$ to Standard Model rate for $H\rightarrow \mu \bar\mu$ as a function of $M_X^2/M_H^2$ for $M_H= 120 $ GeV and $F_X= 250$ GeV with the scale on the vertical axis appropriate for displaying a secondary, much lower peak at $M_X^2/M_H^2 \simeq 0.13$.}
\end{figure}
arises when both of the internal $Z$ lines are basically on-shell thus providing a significant enhancement. For this narrow range of $M_X$ masses which is roughly centered about  $M_X\simeq 13$ GeV, the ratio is  $\sim 10^{-3}$. It then falls quite dramatically by several orders of magnitude as seen in the figure 7. The bump in this  plot arises from one of the internal $Z$ lines being essentially on shell. 
This generic type of behavior is also seen for  a Higgs mass of $180$ GeV as evidenced in the figure 8. Once again, there is a narrow range of $M_X$ values where the decay rate $\Gamma(H\rightarrow \mu\bar\mu XX)$ relative to the Standard Model rate for $H\rightarrow \mu\bar\mu$ is a few parts in $10^{-3}$ while for other $M_X$ values the ratio steeply drops to an experimentally inaccessible range.
\begin{figure}
\includegraphics[width=3.25in]{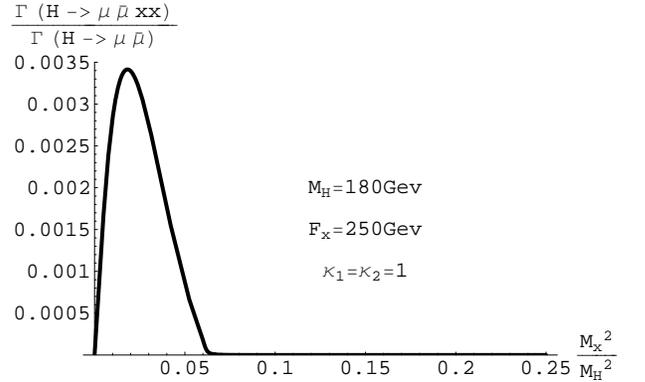}
\caption{Ratio of decay rate for $H\rightarrow \mu\bar\mu XX$ to Standard Model rate for $H\rightarrow \mu \bar\mu$ as a function of $M_X^2/M_H^2$ for $M_H= 180 $ GeV and $F_X= 250$ GeV with the scale on the vertical axis appropriate for displaying the dominant peak.}
\end{figure}

Finally, we consider the decay rate $H\rightarrow \gamma XX$ which appears as  a photon plus missing energy. The leading Feynman graph contributing to this process is displayed in figure 9. 
\begin{figure}
\includegraphics[width=2in]{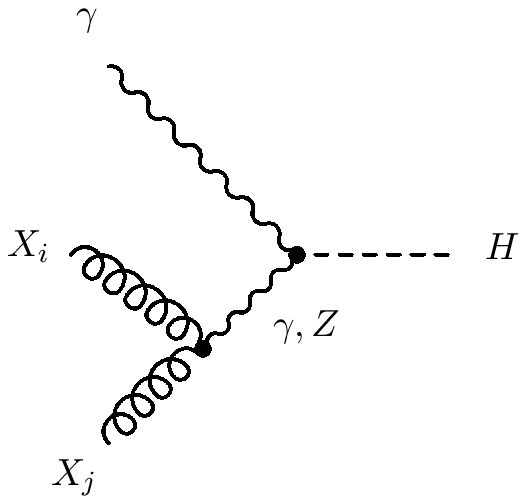}
\caption{Leading order Feynamn graph for the process $H\rightarrow \gamma XX$.}
\end{figure}
Here for the $H\gamma\gamma$ and $H\gamma Z$ vertices, we use the 1-loop Standard Model effective couplings 
\bea
&&S_{H\gamma\gamma}+S_{H\gamma Z}=\frac{\sin^2\theta_W M_W^2}{\sqrt{4\pi M_H^3}}\cr
&&~~~~\int d^4x \biggr[\sqrt{\Gamma_{H\rightarrow \gamma\gamma}} F_{\mu\nu}F^{\mu\nu}H \cr
&&~~~~+ \sqrt{\frac{2\Gamma_{H\rightarrow \gamma Z}}{(1-\frac{M_Z^2}{M_H^2})^3}}F_{\mu\nu}Z^{\mu\nu}H\biggr] ~,
\eea
with $F_{\mu\nu}=\partial_\mu A_\nu-\partial_\nu A_\mu~;~Z_{\mu\nu}=\partial_\mu Z_\nu-\partial_\nu Z_\mu$ which is secured from the fermion and $W$ vector 1-loop  graphs.  On the other hand, the $\gamma XX$ and $ZXX$ vertices are given by the effective Lagrangian ${\cal L}_{BXX}$. The doubly differential decay rate  is 
$\frac{d\Gamma (H\rightarrow \gamma XX)}{dE_\gamma d\Omega_\gamma}$ is isotropic in the Higgs rest frame, while the differential rate with respect to photon energy in this frame takes the form
\bea
&&\frac{d\Gamma (H\rightarrow \gamma XX)}{dE_\gamma}= \frac{N}{384 \pi^3}\frac{M_X^2 M_H^5}{F_X^8} \eta^3(1-\eta)^{\frac{3}{2}}\cr
&&~~~~(1-\eta-\frac{4M_X^2}{M_H^2})^{\frac{3}{2}}\biggr[\kappa_1^2(1-\eta)\cr
&&~~~~+\kappa_2^2(1-\eta -\frac{4M_X^2}{M_H^2})\biggr]\biggr\{\frac{4\cos^2\theta_W}{(1-\eta)^2}\Gamma_{H\rightarrow \gamma\gamma} \cr
&&~~~~ -\frac{4\cos\theta_W \sin\theta_W (1-\eta-\frac{M_Z^2}{M_H^2})}{(1-\eta)[(1-\eta-\frac{M_Z^2}{M_H^2})^2+\frac{M_Z^2\Gamma_Z^2}{M_H^4}]}\cr
&&~~~~\sqrt{\frac{2\Gamma_{H\rightarrow \gamma\gamma}\Gamma_{H\rightarrow \gamma Z}}{(1-\frac{M_Z^2}{M_H^2})^3}}\cr
&&~~~~+\frac{2\sin^2\theta_W}{[(1-\eta-\frac{M_Z^2}{M_H^2})^2+\frac{M_Z^2\Gamma_Z^2}{M_H^4}]}\frac{\Gamma_{H\rightarrow \gamma Z}}{(1-\frac{M_Z^2}{M_H^2})^3}\biggr\} ~,
\eea
where $0\le \eta =\frac{2E_\gamma}{M_H}\le 1-\frac{4M_X^2}{M_H^2}$. In  figure 10, we plot the ratio of the differential rate to the total rate for this process as a function of $\eta=\frac{2E_\gamma}{M_H}$ for $M_X=25$ GeV, while figure 11 shows the ratio of the total rate for this process  to the Standard Model rate for $H\rightarrow \gamma\gamma$.
\begin{figure}
\includegraphics[width=3.25in]{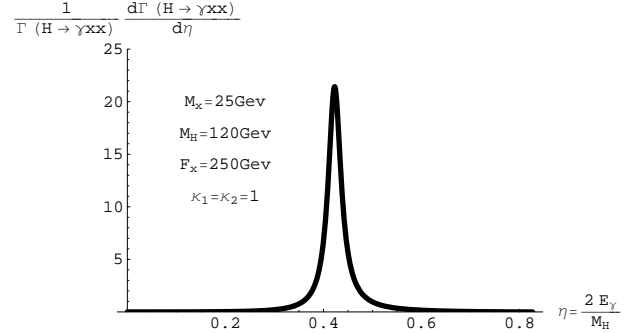}
\caption{Ratio of differential rate $\frac{d\Gamma (H\rightarrow \gamma XX)}{d\eta}$, with $\eta =\frac{2E_\gamma}{M_H}$, to total rate for $H\rightarrow \gamma XX$ as a function of $\eta$.}
\end{figure}
The peak tracks the enhancement accompanying the internal $Z$ line going to mass shell. Unfortunately, the rate for this process is too small to allow for experimental detection.
\begin{figure}
\includegraphics[width=3.25in]{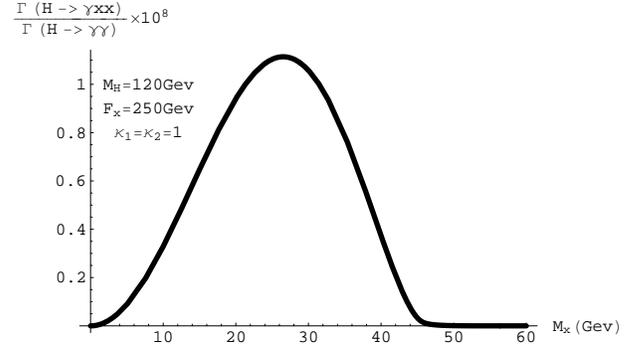}
\caption{Ratio of total rate for $H\rightarrow \gamma XX$ to Standard Model rate for $H\rightarrow \gamma\gamma$ as a function of $M_X$.}
\end{figure}

A variety of flexible brane world models contain  massive, stable vector fields\cite{local},\cite{Kugo}.  In this note we focused on the decay modes of the Higgs boson which contain a pair of such vectors. Using representative values of the vector effective Lagrangian parameters which are consistent with the current experimental limits, we found that the rate for $H\rightarrow XX$ (an invisible Higgs decay) could prove comparable to any of the Standard Model Higgs decay rates. Note that the invisible decay of the Higgs with comparable rates also occurs in other extensions of the Standard Model with additional stable, neutral, weakly interacting particles such as neutralinos in  supersymmetric extensions of the Standard Model\cite{Godbole:2003it}.

The work of TEC, STL and CX  was supported in part by the U.S. Department of Energy under grant DE-FG02-91ER40681 (Task B).  The work of TtV was supported in part by a Cottrell Award from the Research Corporation. TtV would like to thank the theory group at Purdue University for its hospitality during his sabbatical leave.

\newpage

\begin{thebibliography}{100}
\bibitem{local}
T.E. Clark, S. T. Love, M. Nitta, T. ter Veldhuis and C. Xiong,
{Phys. Rev.} D{\bf 75}, 065028 (2007), [arXiv:hep-th/0612147]; S.T. Love,  {J. Phys.} A: { Math. Theor.} {\bf 40}, 7049  (2007), [arXiv:hep-th/0611199]; T.E. Clark, S.T. Love, M. Nitta and T. ter Veldhuis, { Phys. Rev.} D{\bf 72}, 085014 (2005), [arXiv:hep-th/0506094]. 


\bibitem{coset}T.E. Clark, S.T. Love, M. Nitta, T. ter Veldhuis and C. Xiong, {Phys. Rev.} D{\bf 76}, 105014 (2007) [arXiv:hep-th/0703179]; T.E. Clark and S.T. Love, { Phys. Rev.} D{\bf 73}, 025001 (2006), [arXiv:hep-th/0510274]; S.T. Love,  { Mod. Phys. Lett.} A{\bf 20},  2903-2911 (2005), [arXiv/hep-th/0510187]
; T.E. Clark, S.T. Love, M. Nitta and T. ter Veldhuis, { J. Math. Phys.} {\bf 46}, 102304 (2005), [arXiv:hep-th/0501241]
; J. Gomis, K. Kamimura and P. West, { Class. Quant. Grav.}  {\bf 23}, 7369 (2006), [arXiv:hep-th/0607057];
{\it JHEP} {\bf 0610}, 015 (2006), [arXiv:hep-th/0607104]. 




\bibitem{branon}
P. Creminelli and A. Strumia, { Nucl. Phys.} B{\bf 596}, 125 (2001), [arXiv:hep-ph/0007267]; 
A.R. Cembranos, A. Dobado and A.L. Maroto, {J. Phys. A}  {\bf 40}, 6631 (2007),  [arXiv:hep-ph/0611024]. 

\bibitem{phenom}
T.E. Clark, S.T. Love, M. Nitta, T. ter Veldhuis and C. Xiong, "Brane Vector Phenomenology", arXiv:hep-th/0709.4023; S.T. Love, "Vectors and Locally Invariant Brane World Models", to appear in {\bf Proceedings of the 2007 Europhysics High Energy Physics Conference }, Manchester, 2007, {\it J. Phys. CS} (2008).

\bibitem{Kugo}
 M. Bando, T. Kugo, T. Noguchi and K. Yoshioka,
 Phys. Rev. Lett.  {\bf 83}, 3601 (1999) [arXiv:hep-ph/9906549]; 
 T.\ Kugo and K.\ Yoshioka,
  Nucl.\ Phys.\  B {\bf 594}, 301 (2001)
  [arXiv:hep-ph/9912496].

\bibitem{Godbole:2003it}  
R. M. Godbole, M. Guchait, K. Mazumdar, S. Moretti and D. P. Roy,
 Phys. Lett.  B {\bf 571}, 184 (2003),  [arXiv:hep-ph/0304137];  
H. Davoudiasl, T. Han and H. E. Logan,   Phys. Rev.  D {\bf 71}, 115007 (2005),
  [arXiv:hep-ph/0412269]. 




\end{thebibliography}
\end{document}